\documentclass[pre,twocolumn,amsfonts]{revtex4-1}
\usepackage{graphicx,amsmath,amssymb}
\usepackage{tikz,xcolor}
\usepackage{epstopdf}
\usepackage{bm}
\usepackage{mathtools}
\usepackage[caption=false]{subfig}
\definecolor{mypink}{RGB}{255,0,127}
\definecolor{myblue}{RGB}{0,0,255}
\definecolor{mygreen}{RGB}{102,204,0}
\definecolor{myorange}{RGB}{255,128,0}
\definecolor{mypurple}{RGB}{127,0,255}

\begin{document}
%\begin{CJK*}{GBK}{song}
\title{Spontaneous symmetry breaking and discontinuous phase transition for spreading dynamics in multiplex networks}

\author{Ningbo An$^{1}$}

\author{Hanshuang Chen$^{2}$} \email{chenhshf@ahu.edu.cn}

\author{Chuang Ma$^{1}$}

\author{Haifeng Zhang$^{1}$}

\affiliation{$^{1}$School of Mathematical Science, Anhui University, Hefei, 230601, China \\
$^{2}$School of Physics and Materials Science, Anhui University,
Hefei, 230601, China}

\date{\today}

\begin{abstract}
We propose a spreading model in multilayer networks and study the
nature of nonequilibrium phase transition in the model. The model
integrates the susceptible-infected-susceptible (or
susceptible-infected-recovered) spreading dynamics with a biased
diffusion process among different layers. A parameter $\alpha$ is
introduced to control the bias of the diffusion process, such that
each individual prefers to move to one layer with more infected (or
recovered) neighbors for larger values of $\alpha$. Using stochastic
simulations and mean-field theory, we show that the type of phase
transition from a disease-free phase to an endemic phase depends on
the value of $\alpha$. When $\alpha$ is small enough, the system
undergoes a usual continuous phase transition as an effective
spreading rate $\beta$ increases, as in single-layer networks.
Interestingly, when $\alpha$ exceeds a critical value the system
shows either a hybrid two-step phase transition or a one-step
discontinuous phase transition as $\beta$ increases. The former
contains a continuous transition between the disease-free phase and
a low-prevalence endemic phase, and a discontinuous transition
between the low-prevalence endemic phase and a high-prevalence
endemic phase. For the latter, only a discontinuous transition
occurs from the disease-free phase directly to the high-prevalence
endemic phase. Moreover, we show that the discontinuous transition
is always accompanied by a spontaneous symmetry breaking in
occupation probabilities of individuals in each layer.

\end{abstract}
 \maketitle
\section{Introduction}\label{sec1}
Over the past two decades, we have witnessed the power of network
science on modeling dynamical processes in complex systems made of
large numbers of interacting elements
\cite{SIR03000167,PRP06000175,PRP08000093,RevModPhys.87.925,PR.687.1}.
In particular, considerable attention has been devoted to the phase
transitions and critical phenomena in complex networks
\cite{RMP08001275}. Owing to the inherent randomness and
heterogeneity in the interacting patterns, some nontrivial phenomena
have been revealed, such as the anomalous behavior of Ising model
\cite{EPB02000191,PRE02016104,PhysRevLett.104.218701} to a vanishing
percolation threshold \cite{PhysRevLett.85.4626,PRL00005468} and the
absence of epidemic thresholds that separate healthy and endemic
phases \cite{PRL01003200} as well as explosive emergence of phase
transitions
\cite{Science323.1453,PhysRevLett.106.128701,PRP2016(2)}.

However, the vast majority of the previous works considered
dynamical processes in single-layer networks. In fact, many
real-world complex systems are usually composed of interwined
multilayer networks \cite{JCN2014,Bianconi2018}. It has now been
recognized that the study of multilayer networks is fundamental for
enhancing understanding of dynamical processes in networked systems
\cite{PRP2014}. A seminal work from Buldyrev et al.
\cite{Nature2010} showed the catastrophic effect of failure in an
interdependent networks of power grids and computers. Baxter et al.
\cite{PhysRevLett.109.248701} showed that percolation transition in
multiplex networks can be discontinuous and a first-order-like, in
contrast to that in single-layer networks. G\'omez et al.
\cite{PhysRevLett.110.028701} studied a linear diffusion in
multilayer networks and showed that the multiplex structure is able
to speed up the less diffusive of separated layers. The diffusive
properties are related to a structural transition of the multiplex
from a decoupled regime to a systemic regime
\cite{NatPhys8.717.2018}. In particular, some works focused on
spreading processes in multilayer networks
\cite{IEEE2015,NatPhy2.901.2016,Arruda2018physrep}. Cozzo et al.
\cite{PhysRevE.88.050801} showed that the epidemic threshold of the
susceptible-infected-susceptible (SIS) model in a multilayer network
is always smaller than that in any isolated network. Wang et al.
\cite{PhysRevE.88.022801} further showed that the epidemic threshold
can be reduced dramatically when two nodes with dominant eigenvector
components of the adjacency matrices of isolated networks are
linked. Similar results were also obtained by a degree-based
mean-field approach \cite{PhysRevE.86.026106}. However, based on the
percolation theory \cite{J.Stat.Mech.2017.034001}, Dickison et al.
\cite{PhysRevE.85.066109} unveiled one important difference in the
susceptible-infected-recovered (SIR) model with respect to the SIS
dynamics when the coupling between layers is weak. Spreading
processes in structured metapopulations can be well characterized
within the framework of multilayer networks as well
\cite{PhysRevX.8.031039,PhysRevE.87.032809,PhysRevE.90.032806}. In
addition, a variety of dynamics has also been investigated, such as
evolutionary games
\cite{Wang_EPJB88.2015.124,NJP19.073017.2017,NJP20.075005.2018},
synchronization
\cite{PhysRevLett.112.248701,SR6.39033.2016,SA2.e1601679.2016},
opinion formation \cite{PhysRevE.89.062818,NJP18.023010.2016}, and
transportation \cite{PhysRevLett.116.108701,PhysRevLett.120.068301}.

The multiplex structure also provides a convenient framework for
studying the interplay between different dynamical processes.
Granell et al. \cite{PhysRevLett.111.128701} coupled the SIS
spreading taking place on a physical network with another stochastic
Aware-Unaware dynamics in a virtual network. They showed that the
infected agents trigger the mechanism of awareness in the virtual
layer, and thus decreases the incidence of the disease and increases
the epidemic onset. Kan et al. \cite{Kan2017} considered that
susceptible individuals can be informed not only from other aware
individuals through the information network, but also become
self-awareness induced by the infected neighbors in contact network.
They showed that the introduction of the self-awareness can lower
the density of infection, but cannot increase the epidemic
threshold. Wang et al. \cite{SR4.5097.2014} proposed an
asymmetrically interacting bilayer network model to elucidate the
interplay between information diffusion and epidemic spreading, and
they showed that the outbreak of the information can be triggered
not only by its own spreading dynamics but also by the epidemic
outbreak on the counter-layer. Zhang et al.
\cite{PhysRevLett.114.038701} observed a spontaneous explosive
synchronization via adaptively controlling the links in terms of a
local order parameter in single-layer and multilayer networks.
Nicosia et al. \cite{PhysRevLett.118.138302} showed that the
interactions between synchronization and transport dynamics can also
induce a spontaneous explosive synchronization in multiplex
networks.

Very recently, discontinuous phase transition of the
spreading model in multiplex networks has received growing
attention. Vel\'asquez-Rojas and Vazquez \cite{PhysRevE.95.052315}
coupled contact process for disease spreading with the voter model
for opinion formation take place on two layers of networks, and they
showed that a continuous transition in the contact process becomes
discontinuous as the infection probability increases beyond a
threshold. Pires et al. \cite{J.Stat.Mech.2018.053407} proposed an
SIS-like model with an extra vaccinated state, in which individuals
vaccinate with a probability proportional to their opinions.
Meanwhile, individuals update their opinions in terms of peer
influence. They also observed a first-order active-absorbing phase
transition in the model. Jiang and Zhou \cite{Sci.Rep.2018.8.1629}
studied the effect of resource amount on epidemic control in a
modified SIS model on a two-layer network, and they found that the
spreading process goes through a first-order phase transition if the
infection strength between layers is weak. Su et al.
\cite{NJP2018.20.053053} proposed a reversible social contagion
model of community networks that includes the factor of social
reinforcement. They showed that the model exhibits a first-order
phase transition in the spreading dynamics, and that a hysteresis
loop emerges in the system when there is a variety of initially
adopted seeds. Chen et al. \cite{NJP20.013007.2018} studied the
dynamics of the SIS model in social-contact multiplex networks when
the recovery of infected nodes depends on resources from healthy
neighbors in the social layer. They found that as the infection rate
increases the infected density varies smoothly from zero to a finite
small value and then suddenly jumps to a high value, where a
hysteresis phenomenon was also observed.

In this work, we propose a spreading model in multiplex networks and
study, both in simulations and theoretically, the nature of phase
transition from a healthy phase to an endemic phase. In the model,
each individual updates its state via the SIS (or SIR) dynamics and
simultaneously performs a biased diffusion from one layer to the
other depending on the number of infected (or recovered) neighbors
in the target layer. We show that the type of phase transition
depends on the value $\alpha$ of a parameter that controls the
degree of the biased diffusion. For small values of $\alpha$, the
phase transition is customarily continuous or second-order. For
intermediate values of $\alpha$, the phase transition is hybrid. The
system first undergoes a continuous transition from the healthy
phase to a low-prevalence endemic phase and then to a
high-prevalence endemic phase in an abrupt way.  For large values of
$\alpha$, the model solely shows a discontinuous or first-order-like
transition from the healthy phase to the high-prevalence endemic
phase. We attribute the mechanism of the novel discontinuous phase
transition to a spontaneous symmetry breaking in occupation
probabilities of individuals in each layer.

The organization of this paper is as follows. We present the
definition of the model and simulation details in section
\ref{sec2}. Main results for the SIS and SIR dynamics are presented
in section \ref{sec3} and section \ref{sec4}, respectively. We
conclude with the summary of the results in section \ref{sec5}.

\section{Model and Simulation details}\label{sec2}
Let us define our model in multiplex networks as follows. The
network is consisted of $\mathcal {L}$ layers. Each layer contains
the same number of nodes, $N$, and there exists a one-to-one
correspondence between nodes in different layers. The topology in
each layer is described by an adjacency matrix $\textbf{A}^{\ell}$
($\ell=1,\cdots,\mathcal {L}$), whose entries $A_{ij}^{\ell}$ are
defined as $A_{ij}^{\ell}=1$ if there is an edge from node $j$ to
node $i$ in the $\ell$-th layer, and $A_{ij}^{\ell}=0$ otherwise.
Note that the topology at each layer may be different. In time $t$,
each individual $i$ can be located at one of layers, denoted by a
variable $v_i(t)\in\{1, \cdots, \mathcal {L}\}$. The dynamics in
multiplex networks is consisted of an epidemic spreading process and
a diffusion process of individuals among layers. We will use the two
most paradigmatic epidemic models, the SIS model and SIR model. The
two models can be described by the following reactions,
\begin{eqnarray}
S + I\xrightarrow{\lambda }2I,{\kern 1pt} {\kern 10pt}
I\xrightarrow{\mu }S; \label{eq1}
\end{eqnarray}
for the reversible SIS model, while for the irreversible SIR model,
\begin{eqnarray}
S + I\xrightarrow{\lambda }2I,{\kern 1pt} {\kern 10pt}
I\xrightarrow{\mu }R. \label{eq2}
\end{eqnarray}
Here $\lambda$ is the spreading rate that a susceptible individual
is infected when contacting with a single infected individual, and
$\mu$ is the recovering rate that an infected is recovered and turns
to be susceptible again (SIS) or becomes immunized (SIR). For
convenience, we denote the state of each individual $i$ at time $t$
by a variable $\sigma_i(t)\in \{S, I\}$ for the SIS model and
$\sigma_i(t)\in \{S, I, R\}$ for the SIR model. To proceed Monte
Carlo simulation, we discretize the time with a step $\Delta t$ for
the dynamical evolutions. In the subsequent time $t+\Delta t$, all
nodes synchronously update their states and locations according to
the following two steps.

(i) Spreading process. If the individual $i$ is susceptible at time
$t$, $\sigma_i(t)=S$, it becomes infective by contacting with
infected individuals of the same layer. Since the probability that
the individual $i$ is infected by each infective individual of the
same layer is $\lambda \Delta t$, the probability that the
individual $i$ is infected at the time $t+\Delta t$ providing that
(s)he is susceptible at time $t$ is written as
\begin{eqnarray}
P\left( {{\sigma _i}(t + \Delta t) = I\left| {{\sigma _i}(t) = S}
\right.} \right)= \lambda {n_i}(t)  \Delta t,\label{eq3}
\end{eqnarray}
where ${n_i}(t)$ is the number of infective individuals at the same
layer as the individual $i$ at the time $t$. Providing that the
individual $i$ is infective at time $t$, $\sigma_i(t)=I$, on the
other hand, (s)he becomes spontaneously susceptible for the SIS
model (or immunized for the SIR model) at the time $t+\Delta t$ with
a probability $\mu \Delta t$, i.e.,
\begin{eqnarray}
P\left( {{\sigma _i}(t + \Delta t) = S/R \left| {{\sigma _i}(t) = I}
\right.} \right)=\mu \Delta t.\label{eq4}
\end{eqnarray}

\begin{figure}
\centerline{\includegraphics*[width=0.8\columnwidth]{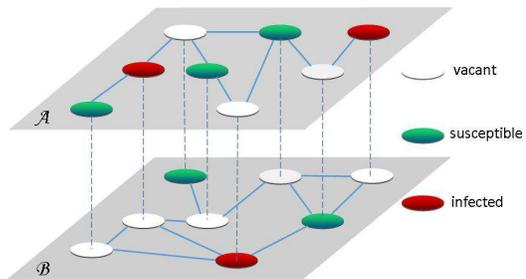}}
\caption{A schematic of our model with the SIS dynamics in a
two-layer network. Each individual stays either in layer $\mathcal
{A}$ or in layer $\mathcal {B}$. If an individual stays in layer
$\mathcal {A}$ at a certain moment, (s)he is vacant in layer
$\mathcal {B}$ at the moment. The state of each individual is either
susceptible (S) or infected (I). The evolution of the states and
locations of all the individuals are governed by the SIS spreading
dynamics (Eq.(\ref{eq3}) and Eq.(\ref{eq4})) and diffusion processes
(Eq.(\ref{eq6})), respectively. \label{fig1}}
\end{figure}

(ii) \emph{Diffusion} process. Each individual $i$ tries to perform
a random diffusion to some layer with the probability $D \Delta t$
with the diffusion rate $D$. Otherwise, the individual $i$ still
stays in its original layer. If a diffusion process happens, a
target layer needs to be chosen in advance. To this end, we suppose
that the probability that the individual $i$ moves to some layer
depends on the number of infected individuals for the SIS model (or
the number of recovered individuals for the SIR model) among its
neighborhoods in such a layer. To formulate this process, we define
the probability that the $\ell$-th layer is selected as the target
layer of the individual $i$ (including the original layer of the
individual $i$) as
\begin{eqnarray}
T_i^\ell \left( t \right) = \left\{ \begin{gathered}
  \frac{{\exp \left[ {\alpha {n_{i,\ell}}\left( t \right)} \right]}}{{\sum\nolimits_{\ell'} {\exp \left[ {\alpha {n_{i,\ell'}}\left( t \right)} \right]} }}, {\kern 5pt} for {\kern 5pt} $SIS$ {\kern 5pt} model  \hfill \\
  \frac{{\exp \left[ {\alpha {n_{r,\ell}}\left( t \right)} \right]}}{{\sum\nolimits_{\ell'} {\exp \left[ {\alpha {n_{r,\ell'}}\left( t \right)} \right]} }}, {\kern 5pt} for {\kern 5pt} $SIR$ {\kern 5pt} model  \hfill \\
\end{gathered}  \right.\label{eq5}
\end{eqnarray}
where the denominator is a normalization factor and $\alpha$ is a
tunable parameter that controls the preference on the choice of the
target layer. $n_{i,\ell} (t)$ ($n_{r,\ell} (t)$) is the number of
infected (recovered) individuals among the neighborhoods of the
individual $i$ in the $\ell$-th layer at the time $t$. If
$\alpha=0$, each layer has an equal probability to be chosen as the
target layer \cite{IEEE2017}. If $\alpha>0$, the individual prefers
to choose the layer with more infected (or recovered) individuals
among its neighborhoods as the target layer for the SIS (or SIR)
model. Instead, if $\alpha<0$ individuals prefer to choose
the layer with less infected or recovered individuals. The larger
the absolute value of $\alpha$ is, the stronger degree of
the preference has. The case $\alpha>0$ may be related to
the spreading dynamics of some virtual information, such as interest
for some product or scientific research topic. While for
the spreading of a real epidemic, one always tries to avoid
infection by moving to layers with the smaller number of infected
individuals, which corresponds to the case $\alpha<0$. Finally, the
probability of the individual $i$ moving to the $\ell$-th layer at
the time $t+\Delta t$ can be written as
\begin{eqnarray}
P\left( {{v_i}(t + \Delta t) = \ell} \right) = \left\{
\begin{gathered}
  1 - D\Delta t+T_i^{\ell}\left( t \right)D\Delta t, {\kern 5pt}   \ell = {v_i}(t); \hfill \\
  T_i^{\ell}\left( t \right)D\Delta t, {\kern 64pt}   \ell \ne {v_i}(t). \hfill \\
\end{gathered}  \right.\label{eq6}
\end{eqnarray}
We also used other forms of Eq.(\ref{eq5}) such as a power-law
function, $T_i^{\ell} \propto n_{i(r),\ell}^\alpha$, and found the
main conclusions shown below hold as well. A schematic of our model
with the SIS dynamics in a two-layer network is shown in
Fig.\ref{fig1}.

\begin{figure}[t]
\centerline{\includegraphics*[width=1.0\columnwidth]{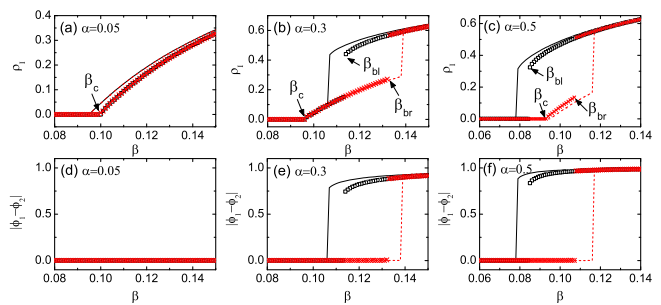}}
\caption{The density of infected individuals $\rho_I$ (top panels)
and $|\phi_1-\phi_2|$ (bottom panels) as functions of the effective
spreading rate $\beta$ in a two-layer network for three distinct
values of $\alpha$: 0.05, 0.3, and 0.5 (from left to right). Each
layer is consisted of an identical ER network with $N=5 \times 10^4$
nodes and average degree $\left\langle k \right\rangle=20$. All the
results are obtained from two different initial conditions:
$\rho(0)=0.1$ and $|\phi_1(0)-\phi_2(0)|=1$ (squares and solid
lines); $\rho(0)=0.1$ and $|\phi_1(0)-\phi_2(0)|=0$ (crosses and
dotted lines). The symbols and lines correspond to simulation and
theoretical results, respectively. \label{fig2}}
\end{figure}

In the simulations, we set $\mu=1$, $D=10$ and $\Delta t =0.05$
unless otherwise specified. We have also tested other sets of
parameters and found that the results are essentially the same. We
define a dimensionless quantity $\beta=\lambda/\mu$ as the effective
spreading parameter, and $\rho_{\{I,R\}} (t)$ as the fraction of the
infected (recovered) individuals at time $t$. Furthermore, we define
${\phi _{\ell}}(t)$ as the occupation probability of individuals in
the $\ell$-th layer at time $t$. Before the simulation, we randomly
choose a fraction $\rho_I(0)$ of individuals as the seeds of the
spreading, i.e., these seeded individuals are initially set to be
infective and the remaining individuals to be susceptible. The
initial locations of all the individuals are prepared with three
distinct cases: (i) $\phi_1(0)=\phi_2(0)=0.5$; (ii) $\phi_1(0)=0$,
$\phi_2(0)=1$, and (iii) $\phi_1(0)=1$, $\phi_2(0)=0$. To achieve
the stationary values of $\rho_{\{I,R\}}$ and $\phi_{\ell}$, we
generate 20 random realizations for a given initial condition. In
each realization, the simulation is run up to $t=10^3$ and the last
$t=5 \times 10^2$ is used to compute the averages.

\section{SIS dynamics}\label{sec3}
We start from a two-layer network. Each layer is consisted of an
identical Erd\"os-R\'enyi (ER) network with $N=5 \times 10^4$ nodes
and average degree $\left\langle k \right\rangle=20$ \cite{ER1960}.
The initial density of infected seeds is set to be $\rho_I(0)=0.1$.
Figure \ref{fig2} shows $\rho_I$ (top panels) and $|\phi_1-\phi_2|$
(bottom panels) as functions of $\beta$ for three distinct values of
$\alpha$: 0.05, 0.3, and 0.5 (from left to right). Since the
topologies in all layers are exactly the same, the initial
conditions (ii) and (iii) produce the same results. Later, we will
discuss the general case when the networks in each layer are not the
same. For $\alpha=0.05$ (Fig.\ref{fig2}(a)), one sees that the
results for two different initial conditions coincide with each
other, implying that the system undergoes a usual continuous
second-order phase transition from a healthy phase (HP) to an
endemic phase (EP) as $\beta$ increases, separated by a threshold
value of $\beta_c$. This is qualitatively the same as the SIS
dynamics in single-layer networks. Quantitatively, the threshold is
raised for the multi-layer case. This is because that all
the individuals stays in each layer with equal probabilities
(Fig.\ref{fig2}(d)) and therefore the bilayer network can be
considered to be two separated percolated networks with site
occupied probability 0.5. In such percolated networks, the degree of
each node is reduced on average by half. It is known that a simple
mean-field calculation predicts that epidemic threshold equals to
the inverse of the average degree of network
\cite{RevModPhys.87.925}. Thus, the value of epidemic threshold in
the bilayer network is expected to be twice that of a single layer
with the same average degree \cite{IEEE2017}. Below we shall show
the conclusion also holds by using an individual-based mean-field
method (see Eq.(\ref{eq15}) and Eq.(\ref{eq16})).

%Note that the SIS model has a unique absorbing phase corresponding
%to no more infected individuals. This implies that the state cannot
%be left once the dynamics brings the system into it.

Interestingly, for $\alpha=0.3$ (Fig.\ref{fig2}(b)), the system
first undergoes a continuous transition from HP to a low-prevalence
endemic phase (L-EP) at $\beta=\beta_c$, and then shows a
discontinuous transition to a high-prevalence endemic phase (H-EP)
at $\beta=\beta_b$. The location of $\beta_b$ depends on the
distribution of initial positions of individuals. When the initial
condition (i) is used, the discontinuous transition point occurs at
$\beta=\beta_{br}$, and at $\beta=\beta_{bl}$ while the initial
condition (ii) or (iii) is used. These two discontinuous transition
points, $\beta_{bl}$ and $\beta_{br}$ ($\beta_{bl}<\beta_{br}$), do
not coincide and thus forms a bistable region, a typical
characteristic of a first-order phase transition. Within this
bistable region, the prevalence is either low or high depending on
the initial condition. Specially, if all the individuals are evenly
placed in each layer initially, the final stationary density of
infected individuals is low. If most of the individuals are placed
on some one layer initially, the density is high. Moveover, we
should note that such a discontinuous transition in $\rho_I$ is
accompanied by a symmetry breaking in occupation probabilities of
individuals in each layer (Fig.\ref{fig2}(e)). Prior to the
discontinuous transition happening, the occupation probabilities are
symmetry, $\phi_1=\phi_2$. Across the discontinuous transition, the
symmetry is broken and the occupation probabilities become seriously
uneven, $\phi_1 \neq \phi_2$.

For $\alpha=0.5$ (Fig.\ref{fig2}(c)), we find that, as compared to
$\alpha=0.3$, the case is similar if the initial condition (i) is
used, but is significantly different if the initial condition (ii)
or (iii) is adopted. For the latter, the system shows the solely
discontinuous transition from the HP to the H-EP at
$\beta=\beta_{bl}$. Thus, we can divide the bistable region into two
subregions. In the left subregion ($\beta_{bl}<\beta<\beta_c$), the
HP and the H-EP coexist (named as bistable phase 1 (BP1)). In the
right one ($\beta_{c}<\beta<\beta_{br})$, the L-EP and the H-EP
coexist (called BP2). Also, the symmetry broken in the occupation
probabilities happens at both boundaries of the bistable region
(Fig.\ref{fig2}(f)).

We have also investigated the case when $\alpha<0$. We
found that the outbreak of the epidemic is always continuous and the
occupation probabilities in each layer are the same. While the
epidemic threshold increases slowly as $\alpha$ decreases, such a
case does not bring any qualitative change in the nature of phase
transition. Therefore, in the following we will focus on the case
only when $\alpha \geq 0$.

The bistability always implies the coexistence of two or
more distinct stable phases within the bistable region. Intuitively
speaking, its origin in our model stems from the the interaction
between spreading dynamics and diffusion in a nonlinear way. To
seek a theoretical interpretation, we define $\rho_{I,i}(t)$ as the
probability that the individual $i$ is infected at time $t$, and
$w_{i,\ell}(t)$ as the probability that the individual $i$ locates
at the $\ell$-th layer at time $t$. The time evolution for
$\rho_{I,i}(t)$ reads,
\begin{eqnarray}
{\rho_{I,i}}(t + \Delta t) = \left[ {1 - {\rho_{I,i}}(t)}
\right]{q_i}(t) \Delta t+ (1-\mu \Delta t) {\rho_{I,i}}(t),
\label{eq7}
\end{eqnarray}
where $q_i(t)$ is the infection rate of the individual $i$ at time
$t$, given by
\begin{eqnarray}
{q_i}\left( t \right) = \lambda \sum\limits_{j = 1}^N
{{\rho_{I,j}}\left( t \right)\sum\limits_{\ell = 1}^{\mathcal{L}}
{A_{ij}^{\ell}} } {w_{i,\ell}}(t){w_{j,\ell}}(t).\label{eq8}
\end{eqnarray}
The time evolution for ${w}_{i,\ell}$ is written as
\begin{eqnarray}
{w_{i,\ell}}\left( {t + \Delta t} \right) = \left( {1 - D\Delta t}
\right){w_{i,\ell}}\left( t \right) + \sum\limits_{\ell'=1
}^{\mathcal {L}} {{w_{i,\ell'}}\left( t \right)} T_i^{\ell}\left( t
\right)D\Delta t.\label{eq9}
\end{eqnarray}
Utilizing the normalization condition $\sum\nolimits_{\ell =
1}^\mathcal {L} {{w_{i,\ell}}(t) = 1}$, Eq.(\ref{eq9}) can be
rewritten as
\begin{eqnarray}
{w_{i,\ell}}\left( {t + \Delta t} \right) = \left( {1 - D\Delta t}
\right){w_{i,\ell}}\left( t \right) + T_i^{\ell} D\Delta t
.\label{eq10}
\end{eqnarray}
In the continuous time limit, $\Delta t \to 0$, Eq.(\ref{eq7}) and
Eq.(\ref{eq10}) can be rewritten as ordinary differential equations,
\begin{eqnarray}
\frac{{d{\rho_{I,i}(t)}}}{{dt}} = \left( {1 - {\rho_{I,i}(t)}}
\right){q_i(t)} - \mu {\rho_{I,i}(t)},\label{eq11}
\end{eqnarray}
and
\begin{eqnarray}
\frac{{d{w_{i,\ell}}(t)}}{{dt}} =  - D{w_{i,\ell}(t)} + D
T_i^{\ell}(t).\label{eq12}
\end{eqnarray}
In the steady state, $d{\rho_{I,i}(t)}/dt=0$ and
$d{w_{i,\ell}(t)}/dt=0$, we have
\begin{eqnarray}
\mu {\rho_{I,i}} = \left( {1 - {\rho_{I,i}}}
\right){q_i},\label{eq13}
\end{eqnarray}
and
\begin{eqnarray}
{w_{i,\ell}} = T_i^{\ell}=\frac{{\exp \left( {\alpha {n_{i,\ell}}}
\right)}}{{\sum\nolimits_{\ell'} {\exp \left( {\alpha {n_{i,\ell'}}}
\right)} }},\label{eq14}
\end{eqnarray}
where ${{ n}_{i,\ell}}$ can be written as ${{ n}_{i,\ell}} =
\sum\nolimits_{j = 1}^N {A_{ij}^{\ell}} {w_{j,\ell}}{\rho_{I,j}}$ by
the mean-field approximation. The stationary solutions of
$\rho_{I,i}$ and $w_{i,\ell}$ can be obtained by numerically
iterating Eq.(\ref{eq13}) and Eq.(\ref{eq14}) together with
Eq.(\ref{eq8}). Once $\rho_{I,i}$ and $w_{i,\ell}$ are
obtained, $\rho_I$ and $\phi_\ell$ can be calculated by ${\rho _I} =
{N^{ - 1}}\sum\nolimits_{i = 1}^N {{\rho _{I,i}}}$ and ${\phi_\ell}
= {N^{ - 1}}\sum\nolimits_{i = 1}^N {{w_{i,\ell}}}$, respectively.
The comparison between the theory and simulations is shown in
Fig.\ref{fig2}. Obviously, the theory can reproduce qualitatively
the main results of the simulations. Quantitatively, the simulations
underestimate the range of the bistable region. One of the main
reasons may be that near the boundaries of the bistable region the
lifetime of one of the metastable states becomes short so that it
cannot be fully sampled in the simulations.

For $\beta$ lower than the epidemic threshold, $\rho_{I,i}=0$, one
notice that $w_{i,\ell}=1/\mathcal{L}$ $\forall i, \ell$ is the only
set solution of Eq.(\ref{eq14}). Substituting
$w_{i,\ell}=1/\mathcal{L}$ into Eq.(\ref{eq13}) and then linearizing
Eq.(\ref{eq13}) around $\rho_{I,i}=0$, we obtain the threshold value
of the spreading rate, $\beta_c$, which is the inverse of the
largest eigenvalue of the matrix $\bar {\textbf{A}}$, defined as
\begin{eqnarray}
\bar {\textbf{A}} = \frac{1}{{{\mathcal{L}^2}}}\sum\limits_{\ell =
1}^\mathcal{L} {{\textbf{A}^{\ell}}},\label{eq15}
\end{eqnarray}
such that
\begin{eqnarray}
{\beta _c} = \frac{1}{{{\Lambda _{\max }}({\bar
{\textbf{A}}})}}.\label{eq16}
\end{eqnarray}

When the topologies of all the layers are the same,
$\textbf{A}^1=\cdots=\textbf{A}^\mathcal{L}=\textbf{A}$, the right
boundary of the bistable region, $\beta_{br}$, can be derived
analytically. Under such a case, it is not hard to check that
$w_{i,\ell}=1/\mathcal{L}$ is always the set of solution of
Eq.(\ref{eq14}). The right boundary of the bistable region is
determined by the condition under which the set of solution loses
its stability. To this end, we need to write down the Jacobian
matrix $\textbf{J}$ of Eq.(\ref{eq12}). Since $w_{i,\ell}$ satisfies
the normalization condition $\sum\nolimits_{\ell} {{w_{i,\ell}} =
1}$, only $\mathcal{L}-1$ variables among $w_{i,\ell}$ ($\ell=
1,\cdots,\mathcal{L}$) are independent of each other. Therefore, we
select $w_{i,1},\cdots,w_{i,\mathcal{L}-1}$ as the independent
variables, and thus $\textbf{J}$ is an
$(\mathcal{L}-1)N$-dimensional square. The entries of $\textbf{J}$
are given by the derivation of right hand side of Eq.(\ref{eq12})
with respective to $w_{i,\ell}$ at $w_{i,\ell}=1/\mathcal{L}$.
Denoting the right hand side of Eq.(\ref{eq12}) by $D
h_{i,\ell}$ with $h_{i,\ell}=- {w_{i,\ell}} + T_i^\ell$, the partial
derivative of $h_{i,\ell}$ with respect to $w_{j,s}$ is given by
\begin{eqnarray}
\frac{{\partial h_{i,\ell}}}{{\partial {w_{j,s}}}} = & -& {\delta
_{ij}}{\delta _{\ell s}} + \frac{{\alpha {A_{ij}}{\rho
_{I,j}}}}{Z}\exp
\left( {\alpha {n_{i,\ell}}} \right){\delta _{\ell s}} \nonumber \\
&-& \frac{{\alpha {A_{ij}}{\rho _{I,j}}}}{{{Z^2}}}\exp \left(
{\alpha {n_{i,\ell}}} \right)\left[ {\exp \left( {\alpha {n_{i,s}}}
\right) - \exp \left( {\alpha {n_{i,\mathcal{L}}}} \right)} \right]
\end{eqnarray}
where $Z = \sum\nolimits_\ell {\exp \left( {\alpha {n_{i,\ell}}}
\right)}$ and $\delta_{ij}$ is the Kronecker symbol defined as
$\delta_{ij}=1$ if $i=j$ and zero otherwise. At
$w_{i,\ell}=1/\mathcal{L}$, ${n_{i,\ell}} = \sum\nolimits_j
{{A_{ij}}} {\rho _{I,j}^*}/\mathcal{L}$ that is independent of the
layer index $\ell$. This leads to
\begin{eqnarray}
{\left. {\frac{{\partial h_{i,\ell}}}{{\partial {w_{j,s}}}}}
\right|_{{w_{i,\ell}} = 1/\mathcal{L}}} =  - {\delta _{ij}}{\delta
_{\ell s}} + \frac{1}{\mathcal{L}}\alpha {A_{ij}}{\rho
_{I,j}^*}{\delta _{\ell s}}
\end{eqnarray}
Therefore, the Jacobian matrix $\textbf{J}$ can be written as the
Kronecker product of two matrices,
\begin{eqnarray}
\textbf{J} =D  \textbf{I} \otimes \textbf{M},\label{eq17}
\end{eqnarray}
where $\textbf{I}$ is an $(\mathcal{L}-1)$-dimensional identity
matrix, and $\textbf{M}$ is an $N$-dimensional square whose entries
are given by
\begin{eqnarray}
{M_{ij}} =  - {\delta _{ij}} + \frac{\alpha}{\mathcal {L}}
{A_{ij}}{\rho ^*_{I,j}},\label{eq18}
\end{eqnarray}
where $\rho_{I,i}^*$ is determined by Eq.(\ref{eq13}) at
$w_{i,\ell}=1/\mathcal {L}$, i.e.,
\begin{eqnarray}
{\rho _{I,i}^*} = \frac{\beta}{\mathcal {L}} \left( {1 - {\rho
_{I,i}^*}} \right)\sum\limits_{j = 1}^N {{A_{ij}}} {\rho
_{I,j}^*}.\label{eq19}
\end{eqnarray}
As mentioned above, $\beta_{br}$ is the point at which the solution
$w_{i,\ell}=1/\mathcal{L}$ loses its stability. This implies that
$\beta_{br}$ is determined whenever the largest eigenvalue of
$\textbf{J}$ is equal to zero, ${\Lambda _{\max }}(\textbf{J})=D
{\Lambda _{\max }}(\textbf{M})=0$. Furthermore, as $\rho_{I,i}^*$ is
bounded between 0 and 1, we have $\Lambda_{\max}(\textbf{M})\leq
-1+\frac{\alpha}{\mathcal {L}} \Lambda_{\max}(\textbf{A})$ according
to Wielandt's Theorem \cite{WielandtTheorem}, and we arrive at a
critical value of $\alpha_c=\mathcal
{L}/{\Lambda_{\max}(\textbf{A})}$ in terms of the condition
${\Lambda _{\max }}(\textbf{M})=0$. Therefore, a bistable phase can
be observed only when $\alpha>\alpha_c$.

\begin{figure}[t]
\centerline{\includegraphics*[width=0.8\columnwidth]{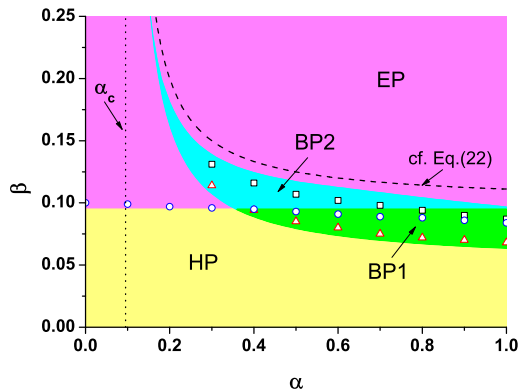}}
\caption{Phase diagram for the SIS dynamics in a two-layer network.
Each layer is consisted of an identical ER network with $N=50000$
and $\left\langle k \right\rangle =20$. Four phases can be
distinguished (encoded in color obtained from the mean-field
theory): the healthy phase (HP), the endemic phase (EP), the
bistable phase 1 (BP1) with the coexistence of the high-density EP
and the HP, and the bistable phase 2 (BP2) with the coexistence of
the high-density and low-density EPs. Vertical dotted line indicates
to the critical value of $\alpha$, $\alpha_c$, giving the lower
bound for observing the bistable phase. Dashed line gives the right
boundary of the bistable phase from the simple mean-field theory
(cf. Eq.(\ref{eq20})). Symbols represents the simulation results for
the boundaries of distinct phases ($\bigcirc$ for epidemic
threshold, $\triangle$ and $\square$ for the left and right boundary
of bistable region, respectively.). \label{fig3}}
\end{figure}

In general, $\rho_{I,i}^*$ and $\beta_{br}$ are obtained only by
numerics. However, in the spirit of a simple mean-field theory we
can derive their analytical results. Let $\rho_{I}^*\equiv
\rho_{I,i}^*$ and $\left\langle k \right\rangle \equiv
\sum\nolimits_j {{A_{ij}}}$, and thus $\rho_{I}^*=0$ or
$\rho_{I}^*=1-\mathcal {L}/( \beta \left\langle k \right\rangle)$ in
terms of Eq.(\ref{eq19}). Substituting the latter solution of
$\rho_{I}^*$ into Eq.(\ref{eq18}) yields ${\Lambda _{\max
}}(\textbf{M}) =-1+ \frac{\alpha }{\mathcal {L}}\left( {1 -
\frac{\mathcal {L}}{{\beta \left\langle k \right\rangle }}}
\right){\Lambda _{\max }}(\textbf{A}) = -1+\frac{\alpha }{\mathcal
{L}}\left( {\left\langle k \right\rangle  - \frac{\mathcal
{L}}{\beta }} \right)$. Note that the simple mean-field theory is at
work when the underlying network is degree-regular random network
for which the largest eigenvalue of the adjacency matrix is
${\Lambda _{\max }}(\textbf{A})=\left\langle k \right\rangle$.
Letting ${\Lambda _{\max }}(\textbf{M})=0$, we obtain an analytical
expression of $\beta_{br}$, given by
\begin{eqnarray}
{\beta _{br}^{SMF}} = \frac{{\mathcal {L}\alpha }}{{\alpha
\left\langle k \right\rangle  - \mathcal {L}}}.\label{eq20}
\end{eqnarray}
Since $\beta_{br}^{SMF}\geq0$ in Eq.(\ref{eq20}), we obtain the
value of $\alpha_c$ in the simple mean-field theory,
$\alpha_{c}^{SMF} =\mathcal {L}/\left\langle k \right\rangle$.
We should note that the asymmetric solution of
Eq.(\ref{eq14}) for $w_{i,\ell}$ can not be obtained analytically,
such that the analytical determination for $\beta_{bl}$ is in
general impossible.

In Fig.\ref{fig3}, we show the phase diagram in $\beta \sim \alpha$
parametric spaces. The whole phase diagram can be divided into three
regions: the HP, the BP, and the EP. The bistable region can be
distinguished with two subregions. For the first bistable phase
(BP1), the H-EP and the HP are coexisting. For the second bistable
phase (BP2), the L-EP and the H-EP are coexisting.

\begin{figure}[t]
\centerline{\includegraphics*[width=1.0\columnwidth]{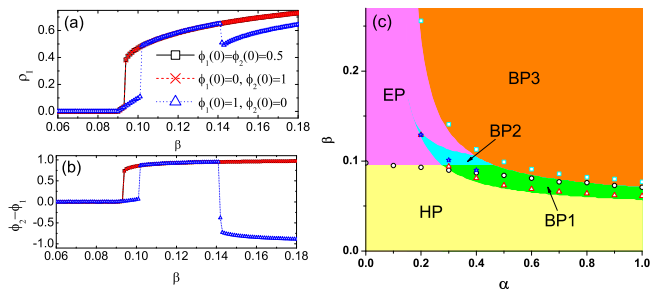}}
\caption{The results for the SIS dynamics on a two-layer network
with distinct topologies in each layer. Each layer is consisted of
an ER network with the same number of nodes $N=5 \times 10^4$ but
different average degrees: $\left\langle k \right\rangle_1 =18$ and
$\left\langle k \right\rangle_2 =22$. The density of infected
individuals $\rho_I$ (a) and the difference in occupying
probabilities of individuals in each layer $\phi_2-\phi_1$ (b) as
functions of $\beta$ at $\alpha=0.3$. (c) shows the phase diagram in
$\beta \sim \alpha$ parametric spaces. Due to topological difference
in each layer an additional bistable region occurs in the
upper-right corner (see Fig.3 for comparison). Symbols represent the
simulation results of the boundaries of distinct phases ($\bigcirc$
for epidemic threshold, and $\triangle$, $\square$, and $\bigstar$
for the boundaries of bistable region). \label{fig4}}
\end{figure}

Finally, we consider the case when the topologies of bilayer
networks are not the same. In Fig.\ref{fig4}(a) and
Fig.\ref{fig4}(b), we show $\rho_I$ and $\phi_1-\phi_2$ as functions
of $\beta$ at $\alpha=0.3$, where each layer is consisted of an ER
networks with $N=5\times 10^4$ nodes but with different average
degrees, $\left\langle k \right\rangle_1 =18$ and $\left\langle k
\right\rangle_2=22$. For $\beta<0.142$, the case is similar to that
in Fig.\ref{fig2} when the topologies of bilayer networks are the
same. In the H-EP, the occupation probability in the second layer is
dominant as we have set the average degree in this layer to be
larger. Interestingly, a new bistable region appears for
$\beta>0.142$, as compared to Fig.\ref{fig2}(b) and
Fig.\ref{fig2}(e). Within the new bistable region (called BP3), two
different H-EPs coexist, corresponding to epidemic spreading widely
either in the first layer or in the second layer. Due to the
inequivalent topologies of bilayer networks, the prevalences are
different. In Fig.\ref{fig4}(c), we show the phase diagram.

\begin{figure}
\centerline{\includegraphics*[width=0.8\columnwidth]{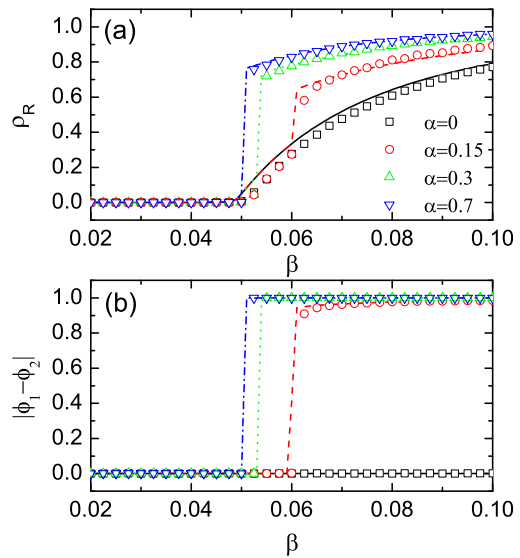}}
\caption{The results for the SIR dynamics. The density of recovered
individuals $\rho_R$ (a) and the absolute value of difference in
occupying probabilities $|\phi_1-\phi_2|$ (b) as functions of
effective spreading rate $\beta$ for four distinct values of
$\alpha$: 0, 0.15, 0.3, and 0.7. Symbols and lines correspond to
simulation and theoretical results, respectively. The multilayer
network is consisted of two identical ER networks with $N=10^6$ and
$\left\langle k \right\rangle =40$. \label{fig5}}
\end{figure}

\section{SIR Dynamics}\label{sec4}
In the SIR model, the number of infected individuals always tends to
zero. The number of recovered individuals reaches only a limited
number of individuals for low spreading rates and a finite fraction
of the population while for high spreading rates. In the
thermodynamic limit, the prevalence $\rho_R$ exhibits a transition
from zero to nonzero as $\beta$ increases. We run the SIR model in a
very large multilayer network consisted of two identical ER random
graphs with $N=10^6$ and $\left\langle k \right\rangle =40$. The
density of initial infection seeds are set to be much small,
$\rho_I(0)=10^{-4}$. We found that the way how to prepare the
initial locations of individuals does not affect the ultimate
prevalence, which may be related to irreversibility of the SIR
model. In Fig.\ref{fig5}(a) and Fig.\ref{fig5}(b), we respectively
show $\rho_R$ and $|\phi_1-\phi_2|$ as functions of $\beta$ for four
distinct values of $\alpha$: 0, 0.15, 0.3, and 0.7. Similar to the
SIS model, for low values of $\alpha$ the model undergoes a usual
continuous phase transition. For large enough values of $\alpha$,
such as $\alpha=0.7$ shown in Fig.\ref{fig5}, the model shows a
discontinuous phase transition. While for the intermediate values of
$\alpha$, such as $\alpha=0.15$ shown in Fig.\ref{fig5}, the model
first shows a continuous transition to a low-density prevalence and
then abruptly jumps to a high-density prevalence. All the
discontinuous transitions are also accompanied by the symmetry
breaking in the occupation probabilities of individuals in each
layer, as shown in Fig.\ref{fig5}(b).

Theoretical analysis can be carried out in a similar way. Let us
denote the probability that individual $i$ is infected (recovered)
at time $t$ by $\rho_{I,i}(t)$ ($\rho_{R,i}(t)$). We can write down
the time evolutions for $\rho_{I,i}$ and $\rho_{R,i}$, given by
\begin{eqnarray}
\frac{{d{\rho_{I,i}(t)}}}{{dt}} = \left( {1 - {\rho_{I,i}(t)}}
-{\rho_{R,i}(t)} \right){q_i(t)} - \mu {\rho_{I,i}(t)},\label{eq21}
\end{eqnarray}
and
\begin{eqnarray}
\frac{{d{\rho_{R,i}(t)}}}{{dt}} =  \mu {\rho_{I,i}(t)},\label{eq22}
\end{eqnarray}
where $q_i(t)$ is also given by Eq.(\ref{eq8}). The occupation
probabilities of each individual in each layer, $w_{i,\ell}(t)$, are
governed by Eq.(\ref{eq12}) as well. Near phase transition,
$\rho_{R,i}$ is macroscopically infinitesimal, and thus
Eq.(\ref{eq21}) recovers to Eq.(\ref{eq11}). Therefore, the
threshold of the SIR model is also determined by Eq.(\ref{eq16}). We
numerically integrate Eqs.(\ref{eq21},\ref{eq22}) and
Eq.(\ref{eq12}), and obtain the theoretical results, as indicated by
lines in Fig.\ref{fig5}. There are in excellent agreement between
simulations and theory.

\section{Conclusions}\label{sec5}
In conclusion, we have proposed a spreading model in multilayer
networks, and studied the nature of nonequilibrium phase transition
as the effective spreading parameter varies. Our model is composed
of a spreading process and a biased diffusion process between
different layers. Using the SIS and SIR models as two paradigmatic
examples of spreading dynamics, we found that the model in
multilayer networks exhibits more abundant behaviors of phase
transition than that in single-layer networks. For the SIS dynamics,
when a biased parameter above a critical value the transition can be
either continuous from the healthy phase to the low-prevalence
endemic phase and then be discontinuous to the high-prevalence
endemic phase, or be discontinuous directly from the healthy phase
phase to the high-prevalence endemic phase, depending on the value
of such a biased parameter and the distribution of initial positions
of individuals. Between the two discontinuous transition points, the
system is bistable and the occupation probabilities of individuals
in each layer show a spontaneous symmetry breaking across these
discontinuous transition points. For the SIR dynamics, we found that
the nature of phase transition is also essentially changed as the
biased parameter exceeds a critical value, similar to the case of
SIS dynamics. However, we did not observe the bistability for the
irreversible SIR model. Furthermore, we have developed an
individual-based mean-field theory that can fully describe the
nature of phase transitions in multilayer networks observed in
simulations. Since the biased diffusion of individuals can lead to
an abrupt explosion of epidemic, it will certainly provide a
challenge as how to predict and control epidemic in real networks
usually represented by multilayer networks
\cite{wang2016statistical}.

\begin{acknowledgments}
We acknowledge the supports from the National Natural Science
Foundation of China (Grants  No. 61473001, No. 11875069) and the
Natural Science Foundation of Anhui Province (Grant No.
1808085MF201).
\end{acknowledgments}

\end{document}